\documentclass[aps,twocolumn,pra,superscript,floatfix,superscriptaddress,showpacs,footinbib]{revtex4-1}
\usepackage{amssymb}

\usepackage{epsf}
\usepackage{epsfig}
\usepackage{graphicx}
\usepackage{dcolumn}
\usepackage{braket}
\usepackage{bm}
\usepackage{amsfonts}
\usepackage{amsmath}
\usepackage{amssymb}
\usepackage{color,soul}
\usepackage{wasysym}
\usepackage{mathrsfs}
\usepackage{natbib}
\usepackage{multirow}
\usepackage[colorlinks=true,%
            linkcolor=blue,%
            urlcolor=blue,%
            citecolor=blue,%
            filecolor=blue,%
            bookmarksopen=true,%
            pdfauthor={J. P. Ramos-Andrade},%https://www.overleaf.com/10769193tdpqskjydmcs
            pdftitle={QD_c_MBS},%
            pdfsubject={QD_c_MBS_Manuscript},%
            pdfpagemode=UseOutlines]{hyperref}

\newcommand{\ve}{\varepsilon}

\newcommand{\bea}{\begin{eqnarray}}
\newcommand{\eea}{\end{eqnarray}}

\newcommand{\up}{\uparrow}
\newcommand{\dn}{\downarrow}

\newcommand{\la}{\langle}
\newcommand{\ra}{\rangle}

\begin{document}

\title{Majorana bound state in the continuum: Coupling between Majorana bound state and quantum dot mediated by
continuum}
\author{J. P. Ramos-Andrade}
\email{juan.ramosa@usm.cl}	
\affiliation{Departamento de F\'isica,  Universidad T\'ecnica Federico
Santa Mar\'ia, Casilla 110 V, Valpara\'iso, Chile}
\affiliation{Instituto de F\'isica,  Universidade Federal de Uberl\^andia,
Uberl\^andia, Minas Gerais 38400-902, Brazil}
\author{P. A. Orellana}
\affiliation{Departamento de F\'isica,  Universidad T\'ecnica Federico
Santa Mar\'ia, Casilla 110 V, Valpara\'iso, Chile}
\author{E. Vernek}
\affiliation{Instituto de F\'isica,  Universidade Federal de Uberl\^andia,
Uberl\^andia, Minas Gerais 38400-902, Brazil}
\affiliation{ Department of Physics and Astronomy, and Nanoscale and Quantum Phenomena Institute, Ohio University, Athens, Ohio 45701–2979, USA}
\begin{abstract}

In this work, we consider a single-level quantum dot (QD) and a Majorana bound state (MBS) placed at the end of a topological superconducting nanowire (TSW). Both are coupled to the continuum and do not have a direct connection between them. We addressed the behavior of MBS leaking phenomena and its consequences into the QD physics in non-interacting and Coulomb blockade regime. By employing Green's function formalism via the equation of motion
procedure, we calculate the physical quantities of interest. Our results
show that the leakage of the MBS into the continuum state is achieved and
can alter the physics of Coulomb blockade in the system through
continuum-mediated coupling between MBS and QD. As a main consequence, we found a robust and non-trivial mechanism to accomplish a bound state in the continuum in the system.

\end{abstract}

\date{\today}
\maketitle

\section{Introduction}
The progress of theoretical formulations and experimental techniques in condensed matter physics provides an interesting playground for scientists to investigate at low energy intriguing phenomena, commonly associated to elementary particles that would be possible solely in high energy physics~\cite{wilczek2009majorana}. An example is the possibility to observe Majorana bound states (MBSs) that have similarities with  Majorana fermions, predicted to exist as elementary particles~\cite{majorana1937}. In condensed matter, MBSs are predicted to emerge as collective excitations in $p$-wave topological superconductors (TSC)~\cite{PhysRevB.61.10267,Ivanov2001,Fu2008}. The MBSs satisfy non-Abelian statistics and have a great deal excitement towards applications in quantum computation~\cite{KITAEV20032,prb.82.020509,alicea2011non,alicea2016superconductors}. After the theoretical proposal performed by Kitaev~\cite{kitaev2001}, in which MBSs would emerge
bound to edges of a one-dimensional (1D) TSC~\cite{PhysRevB.81.125318}, several experiments have been
carried out the physical realization of Kitaev model, finding signatures of
their presence through anomalies in physical quantities measurement~\cite{mourik2012signatures,deng2012anomalous,das2012zero,Churchill2013,albrecht2016exponential,deng2016majorana,Lee2012}. However, from these early physical realizations to practical use, a long way still has to be paved. For example, a full understanding of the transport properties of MBS-based systems are mandatory to use them as electronic devices~\cite{PhysRevB.86.094503}.

Owing to the great flexibility to control their electronic properties, the quantum dots (QDs) have proven to be a convenient platform to study MBS in condensed matter~\cite{Cifuentes2019,RamosAndrade2019,RamosAndrade2018,Schuray2017,Liu2017,Silva2016}.
In a first attempt along these ideas, an attracting system was proposed Liu
and Baranger~\cite{liu2011}. In their proposal, a QD was coupled
simultaneously to two  normal metallic contacts and to the end of a
topological quantum wire holding a MBS. From a theoretical point of view
--and with potential practical application-- an interesting feature noted
by these authors was the half-integer conductance between the normal
contact across the QD.
Later on, it was showed that this  characteristic is obtained
regardless the QD energy level, describing it as a MBS leaking phenomena
~\cite{vernek2014}. In fact this leaking phenomena was observed
experimentally~\cite{deng2016majorana}.

In the setup discussed above, the Majorana mode leaked into the QD because they were directly coupled to each other. The reader may ask what would
happen if the Majorana mode were not directly coupled to the dot. In particular, if a continuum of states mediated the coupling between the MBS and the QD. Our results show that no matter how strong the MBS is coupled to the continuum, its bound state character remains unchanged
\footnote{Note that although QD is not directly connected to the MBS it is close enough so that the local nature of the  Majorana wave function is still probed~\cite{PhysRevB.86.085408}
}. As such, as far as the MBS {\it plus} the contacts concern,  this problem can be viewed as a bound state in the continuum (BIC) akin to the prediction by  von Neumann and Wigner  in a generic framework of engineered potential~\cite{vonNeumann1929} and later investigated in many fermion systems (see Ref.~\cite{Hsu2016} and references therein).   However, this bound state does not represent a full fermion, as in the traditional case, it rather corresponds to an MBS or
"half fermion"; as it is commonly referred. Hence we refer to this state as Majorana bound state in the continuum (MBIC). Recently, BICs has gained considerable attention as it has been observed in
photonic systems. Motivated by the interference phenomena taking place in
electronic systems in analogy with the photonic counterpart, the presence
of BICs promoted by MBS has been investigated~\cite{Ramos2014,DeGuevara2006}. Related to this problem, interplay between
MBSs and BICs have been proposed as a useful tool to perform applications
in quantum computing, allowing, for instance, to read/write information through veil/unveil these
states~\cite{ricco2016decay,Guessi2017,zambrano2018}. Indeed, MBS provides
a quite attractive way to produce  BICs as they are topologically protected against local perturbation~\cite{prb.85.165124,PhysRevB.98.085125}. As a result, manipulating electronic properties of QDs becomes much suitable as the rest of the system turns out to be almost insensitive to applied electric fields.

\begin{figure}[h!]
\centering
 \includegraphics[width=\linewidth]{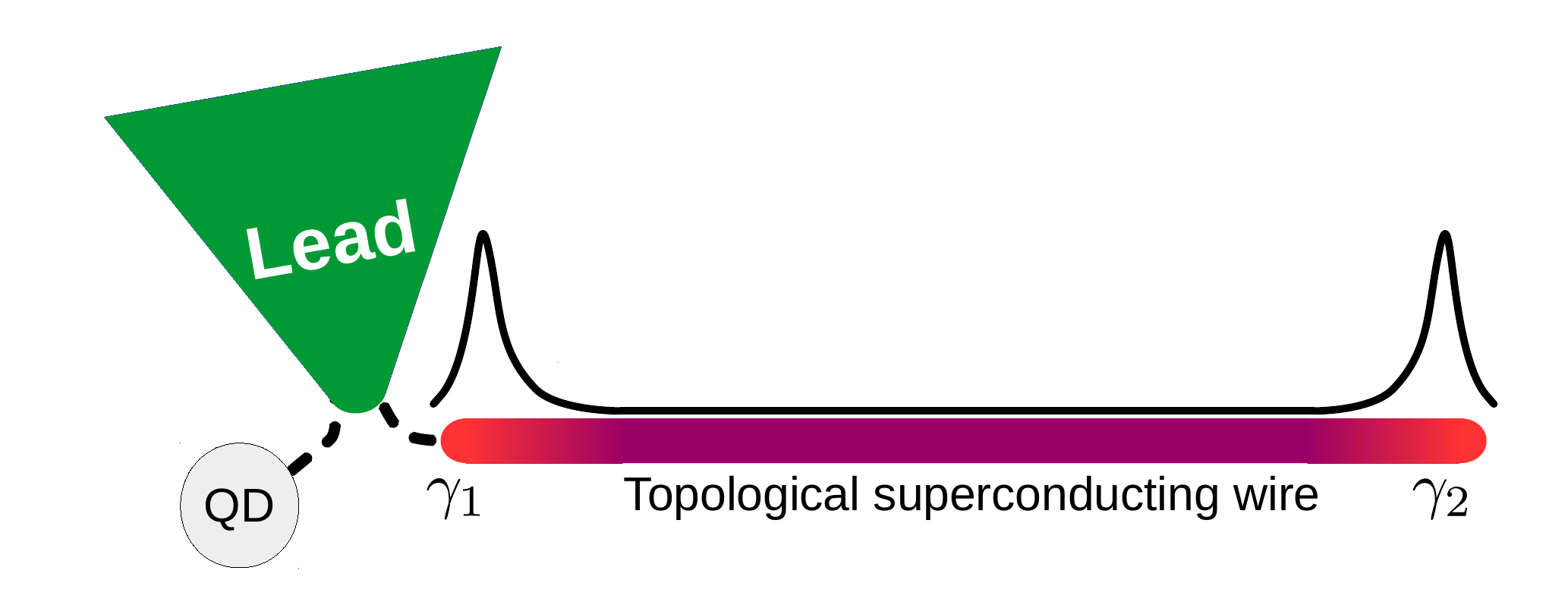}
\caption{Schematic representation of model: A single level QD (gray) and a TSW, hosting MBSs (orange) $\gamma_1$ and $\gamma_2$, coupled with a common metallic lead (green) with a continuum spectrum. The black curve above the topological wire intends to represent the wave function associated to the Majorana bound states (note the peaks at the ends).}
\label{fig1}
\end{figure}

In this work, we propose to study the electronic properties of a system composed of a QD and a topological superconducting wire (TSW), both connected to a common metallic contact.
The TSW is assumed to be in its topological phase,
holding MBSs in its ends. The system is schematically depicted in
Fig.~\ref{fig1}. Alternatively, this system can be viewed as a QD coupled
to an effective continuum exhibiting an MBSs. By employing the Green's
function method the equation of motion techniques, we study the spectral
and transport properties of the system. While in the non-interacting regime
of the QD, we can access the physical property exactly, in its interacting regime, they are available only under certain approximation. Here we employ
the so-called Hubbard I approximation that is known to capture qualitatively well the many-body physics in Coulomb blockade regime~\cite{J.Phys.F:MetalPhys..11.2389}.
{Our results show that no matter how strong the MBS is coupled to the continuum, it features in the QD spectral function as a bound state. This behavior remains unchanged in the strong Coulomb interaction regime of the QD.}

This paper is  organized as follows: Section~\ref{Model} presents the
system Hamiltonian and method used to obtain quantities of interest;
Section~\ref{Results} shows the corresponding results and the related
discussion.  Finally, our concluding remarks are presented in
Section~\ref{Conclu}.

\section{Hamiltonian model and method}\label{Model}
For the sake of  completeness, the system under study consists of a QD and
a MBS located at the end of a TSW, both connected with a common normal
metallic lead, as schematically shown in Fig.~\ref{fig1}. The Hamiltonian
of the system can be written as
\begin{equation}
H=H_{\text{c}}+H_{\text{dot}}+H_{\text{c-dot}}+H_{\text{c-MBS}}\,,\label{H}
\end{equation}
where the first three terms of Eq.~\eqref{H} correspond to the
traditional Anderson Hamiltonian describing the QD plus the normal metallic
lead and are given by
\begin{eqnarray}
H_{\text{c}}&=& \sum_{{\bf k},\sigma}\ve_{{\bf k}}c_{{\bf
k},\sigma}^{\dag}c_{{\bf k},\sigma}\,,\\
H_{\text{dot}}&=&\sum_{\sigma}\ve_{d}d_{\sigma}^{\dag}d_{\sigma}+Un_{
\uparrow}n_{\downarrow}\,,\\
H_{\text{c-dot}}&=&\sum_{{\bf k},\sigma}\left(V_{{\bf k}}c_{{\bf k},
\sigma}^{\dag}d_{\sigma}+V_{{\bf k}}^{\ast}d_{\sigma}^{\dag}c_{{\bf
k},\sigma}\right)\,,
\end{eqnarray}
where $c_{{\bf k},\sigma}^{\dag}(c_{{\bf k},\sigma})$ creates(annihilates)
a continuum electron with momentum ${\bf k}$ and spin $\sigma$;
$d_{\sigma}^{\dag}(d_{\sigma})$ does it in the QD with energy level
$\ve_{d}$, $U$ is the electron-electron interaction, $n_{\sigma}=
d_{\sigma}^{\dag} d_{\sigma}$  is the number operator; and $V_{{\bf k}}$
represents the tunneling matrix element between the continuum states and
the QD orbitals. The last term in Eq.~\eqref{H}, describes the coupling
between the MBS and the lead  and is given by
\begin{equation}\label{HcMBS}
H_{\text{c-MBS}}=\sum_{{\bf k}}\lambda\left(c_{{\bf k},\downarrow}- c_{{\bf
k},\downarrow}^{\dag}\right)\gamma_{1}\,,
\end{equation}
in which $\lambda$ represents the coupling parameter and $\gamma_{1}$  the
MBS operator, which satisfy both $\gamma_{1}^{\dag}=\gamma_{1}$ and
$\gamma_{1}^{2}=1$. Here we have assumed that the Majorana mode is provided
by a long TSW that are fully polarized with spin
{\it down} by an effective magnetic field along the $z$ direction. Hence
only electrons with spin {\it down} couples to the MBS. Moreover, it is
worth of mentioning that we consider a TSW in long-wire limit, then the MBS
placed at the opposite end, $\gamma_{2}$, is strictly equivalent to
$\gamma_{1}$.

We are interested in to study the influence of the MBS onto the
physical properties of the QD, mediated by the continuum. To access the
relevant physical quantities we employ the Green's function formalism which
allows us to obtain, for instance the spin-resolved local density of states
(LDOS) at the QD, $\rho_{d,\sigma}(\ve)$ and transport properties. In
terms of the Green's function (GF), the spin dependent LDOS is given
\begin{equation}\label{LDOS}
\rho_{d,\sigma}(\ve)=-\frac{1}{\pi}\text{Im}\left[\la\la d_{\sigma},d_{\sigma}^{\dag}\ra\ra_{\ve}\right]\,,
\end{equation}
where $\la\la d_{\sigma},d_{\sigma}^{\dag}\ra\ra_{\ve}$ denotes the
spin-resolved retarded GF of the QD in energy domain. In the following, we will address the model either in the non-interacting ($U = 0$) and
interacting ($U > 0$) cases. For $U > 0$ it is known that one cannot obtain
an exact expression for the Green's function. However, approximated
versions can still be obtained. For instance, the so-called Hubbard I
approximation is known to provide a fairly good description of Coulomb blockade phenomena above the Kondo temperature. Such an expression can be
derived by using the equation-of-motion technique, as discussed in
Appendix \ref{appa}. Within this approximation, the Green's function
acquires the form
\begin{align}\label{Self_G}
&\la\la d_{\sigma},d_{\sigma}^{\dag}\ra\ra_{\ve}=\\&\frac{\ve-\ve_{d}
-U(1-\la n_{\bar{\sigma}}\ra)}{(\ve-\ve_{d})(\ve-
\ve_{d}-U)-(\ve-\ve_{d}-U(1-\la
n_{\bar{\sigma}}\ra))\Sigma_{\sigma}(\ve)}\,,\nonumber
\end{align}
in which
\begin{subequations}
\begin{eqnarray}\label{SigmaUP}
\Sigma_{\uparrow}(\ve)&=&-i\Gamma\\\label{SigmaDN}
\Sigma_{\downarrow}(\ve)&=&-i\frac{\Gamma}{1-M(\ve)}
\end{eqnarray}
\end{subequations}
are  the spin-resolved self-energies of the QD. In the
above,  $\la n_{\sigma}\ra$ is the occupation of the QD for a given spin
$\sigma$ and $\Gamma=\left(\pi V^{2}/2D\right) \Theta(D-|\omega|)$ (with
$D$ being the bandwidth of the metallic contact) represents the
energy-independent hybridization parameter between the continuum and QD,
which is derived in the wide-band limit ($D$ much larger than any other
energy parameter of the system). Note that $\Sigma_{\dn}(\ve)$ is modified
by the presence of MBS accounted by the function $M(\ve)$, which is given
by (see Appendix\ \ref{appa})
%\begin{equation}
%
\begin{align}
&M(\ve)=-2i\Lambda\times\\&\left[\ve+\frac{2i\Lambda(\ve+\ve_{d})(\ve+
\ve_{d}+U)}{(\ve+\ve_{d})(\ve+\ve_{d}+U)+i\Gamma(\ve+\ve_{d}+U(1-\la
n_{\uparrow}\ra))}\right]^{-1}\,,\nonumber
\end{align}
%\end{equation}
%
where $\Lambda=\pi\lambda^{2}/2D$ is the hybridization strength between the
 MBS and the continuum. Since the GF~\eqref{Self_G} depends on the
occupation, it must be determined self-consistently.

\section{Numerical Results}\label{Results}

To show our numerical results let us set the hybridization $\Gamma$  as the
energy unit.
In the following, we will analyze the LDOS as a function of the energy for
different values of the relevant parameters of the system, e. g. $\Lambda$,
$\ve_{d}$ and $U$. We shall  first discuss the result at the
non-interacting case, $U = 0$ and next; we will address the case of $U\neq
0$.

\subsection{Non-interacting quantum dot ($U=0$)}\label{subsecU0}

For $U = 0$, the expression for the GF \eqref{Self_G} becomes exact and
acquires the form
\begin{equation}\label{GFU0}
\la\la d_{\sigma};d_{\sigma}^{\dag}\ra\ra_{\ve}=\frac{1}{\ve-\ve_{d}-
\Sigma_{\sigma}(\ve)}\,.
\end{equation}
The effect of the MBS in the QD is accounted by the self-energy
$\Sigma_{\dn}$, via
\begin{equation}\label{MU0}
M(\ve,U=0)=-2i\Lambda\left[\ve+2i\Lambda\frac{\ve+\ve_{d}}{\ve+\ve_{d}
+i\Gamma}\right]^{-1}.
\end{equation}
Note that since the electron's spins are decoupled from each other,  the
spin  $\up$, component is not affected by the MBS. Therefore, we focus only
on the electrons with spin $\dn$ in the QD. As usual, the self-energy
encompasses the information from the rest of the system by shifting the
energy level of the QD by an amount $\text{Re}[\Sigma_{\dn}(\ve)]$ and
broadening the bare level by a quantity $-\text{Im}[\Sigma_{\dn}(\ve)]
\equiv\Gamma_{\text{eff}}(\ve)$. The latter represents the effective
hybridization between the QD and continuum, modified by the MBS. For
$\lambda\rightarrow 0$, $\Gamma_{\text{eff}}\rightarrow\Gamma$, that is
independent of $\ve$.

\begin{figure}[tbph]
\centering
\includegraphics[width=0.99\linewidth]{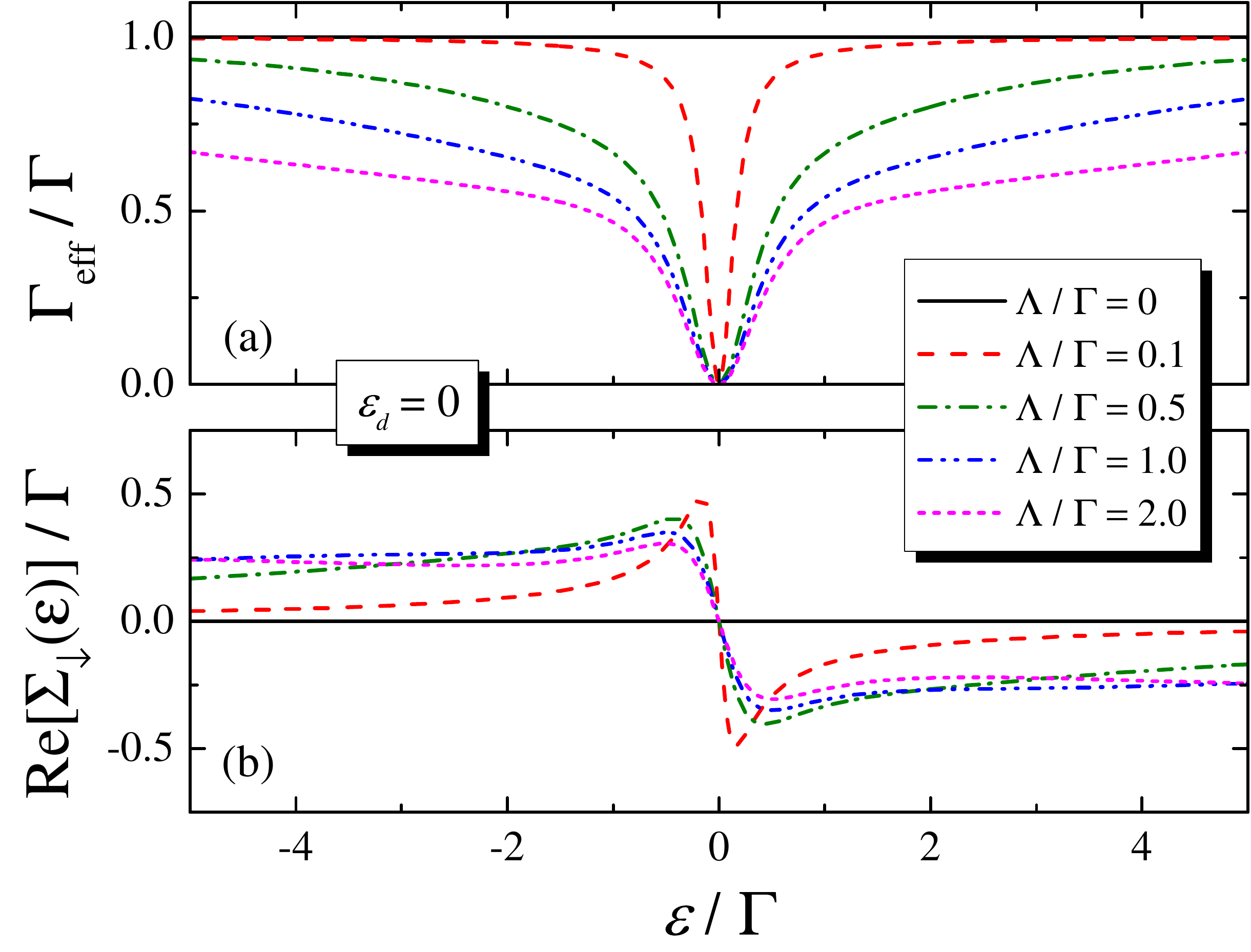}
\caption{(a) $\Gamma_{\text{eff}}$ and (b)  $\text{Re}[\Sigma_{\dn}(\ve)]$
as function of the energy for different MBS-lead couplings $\Lambda$. Here the QD
energy level is fixed at $\ve_{d}=0$. The flat black curve for $\Lambda=0$ reflects the wide band limit assumed in the calculations.}
\label{fig2}
\end{figure}

\begin{figure}[tbph]
\centering
\includegraphics[width=0.99\linewidth]{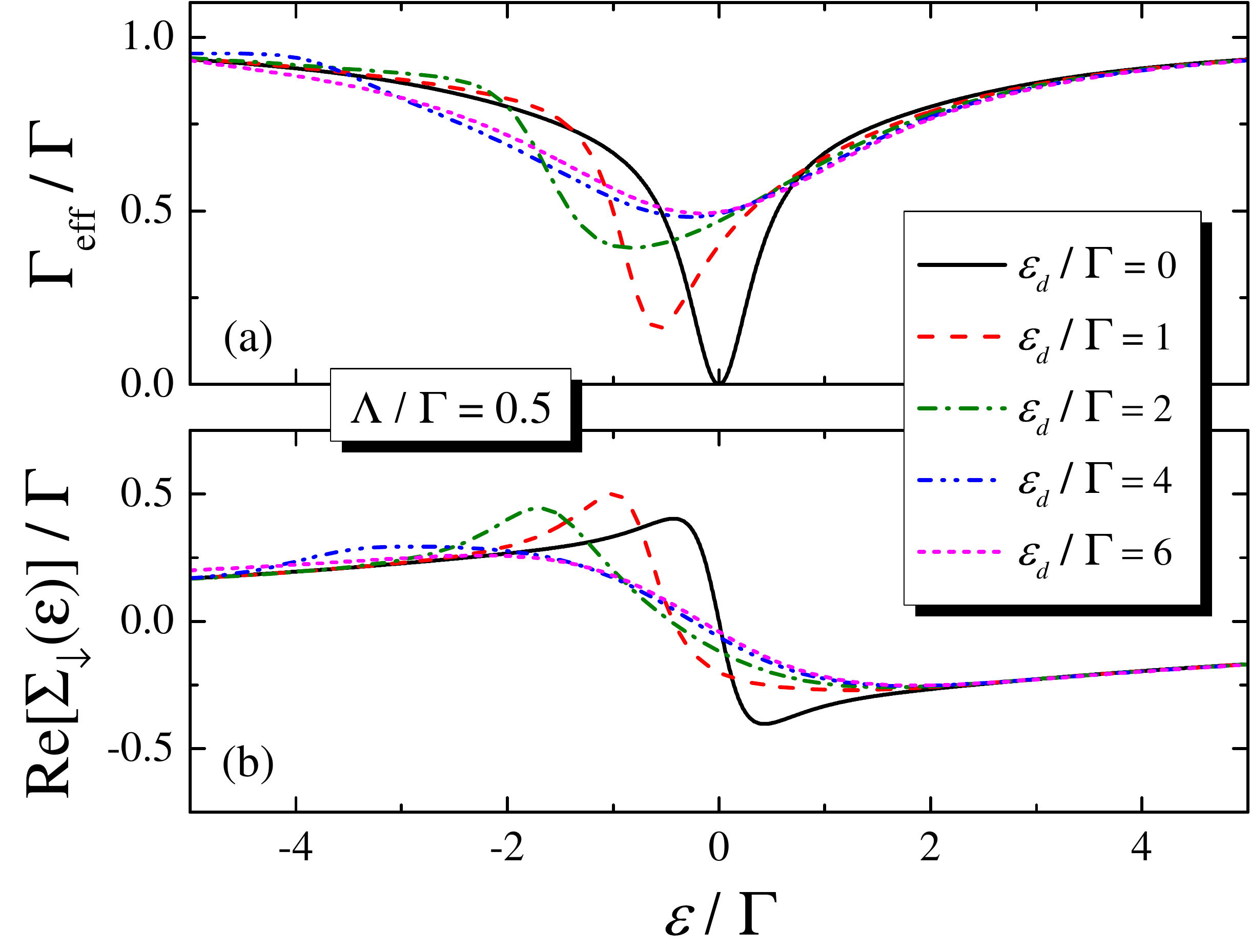}
\caption{(a) $\Gamma_{\text{eff}}$ and (b)  $\text{Re}[\Sigma_{\dn}(\ve)]$
as function of the energy for different QD energy levels $\ve_{d}$. Here the
MBS-continuum coupling is fixed at $\Lambda/\Gamma=0.5$. Note in (a) that $\Gamma_{\rm eff}$ vanishes only for $\ve_d=0$.}
\label{fig3}
\end{figure}

In Fig.~\ref{fig2} we show $\text{Re}[\Sigma_{\dn}(\varepsilon)]$ and
$\Gamma_{\text{eff}}(\varepsilon)$ as a function of $\ve$, using $\ve_{d}=0$
and different values of $\Lambda$. First of all, it is interesting noting
in Fig.~\ref{fig2}(a) that $\Gamma_{\text{eff}} (\varepsilon=0)=0$ for any
value of $\Lambda\neq 0$. This is somewhat surprising because it results
from a destructive quantum interference ---involving a ``half" fermion---
and is very much similar to the case of a  conventional fermion in the
continuum.  This  complete antiresonance at $\ve=0$ decouples the electrons
with spin $\dn$ of the QD from the continuum.
For $|\ve|\gg \Gamma$ we note that $\Gamma_{\rm eff.}$ tends to saturate
at different values depending on how big is $\Lambda$. This can be
understood analytically.  In the limit $\Lambda/\Gamma\gg 1$ and
$\ve_{d}=0$, Eq.~\eqref{MU0} becomes
\begin{equation}
\Gamma_{\text{eff}}(\varepsilon,\varepsilon_{d}=0)=\frac{\Gamma}{2}\frac{
\varepsilon^{2}}{\varepsilon^{2}+(\Gamma/2)^{2}}~,
\end{equation}
which is independent of $\Lambda$. From this equation it is easy to see
that for energies $|\ve|\gg\Gamma$ we obtain
$\Gamma_{\text{eff}}=\Gamma/2$.

 In Fig.~\ref{fig2}(b) we show the real part of the self energy. Note that,
by virtue of the wide-band limit, $\text{Re}[\Sigma_{\dn}(\varepsilon)]=0$
for $\Lambda=0$. Moreover, $\text{Re}[\Sigma_{\dn}(\varepsilon=
\varepsilon_{d} =0)]=0$  for any value of $\Lambda$  and becomes finite for
$\varepsilon\neq 0$, but restrict to the  condition
$|\text{Re}[\Sigma_{\dn}(\varepsilon)]| <\Gamma/2$.

In Fig.~\ref{fig3} we show $\Sigma_{\dn}(\varepsilon)$ for fixed $\Lambda$
and different $\varepsilon_{d}>0$. In Fig.~\ref{fig3}(a) we see that
$\Gamma_{\rm eff}$ vanishes only for $\varepsilon_{d}=0$. Moreover, we note
that $\Gamma_{\text{eff}} (\varepsilon=0)\rightarrow\Gamma/2$ for large
$\varepsilon_{d}$. This is a a remarkable signature of the presence of the
Majorana zero mode in the continuum. In the limit $\ve_{d}\gg\Gamma$, the
contribution given by the MBS to $\Gamma_{\text{eff}}$ amounts to $M(\ve;U
= 0) = 2i\Lambda[\ve+2i\Lambda]^{-1}$. With this we obtain
\begin{equation}
\Gamma_{\text{eff}}(\ve,\ve_{d}\gg\Gamma)=\Gamma\left(\frac{\varepsilon^{2}}{\varepsilon^{2}+16\Lambda^{2}}+\frac{8\Lambda^{2}}{\varepsilon^{2}+16\Lambda^{2}}\right).
\end{equation}
This clearly show that $\Gamma_{\text{eff}}=\Gamma/2$  as $\ve\rightarrow
0$ regardless the value of $\Lambda$. Interestingly, similar to what was
observed in Fig.~\ref{fig2}(b), in Fig.~\ref{fig3}(b)
$\text{Re}[\Sigma_{\dn} (\varepsilon)]$  is also limited as
$|\text{Re}[\Sigma_{\dn} (\varepsilon)]| <\Gamma/2$.

\begin{figure}[tbph]
\centering
\includegraphics[width=0.99\linewidth]{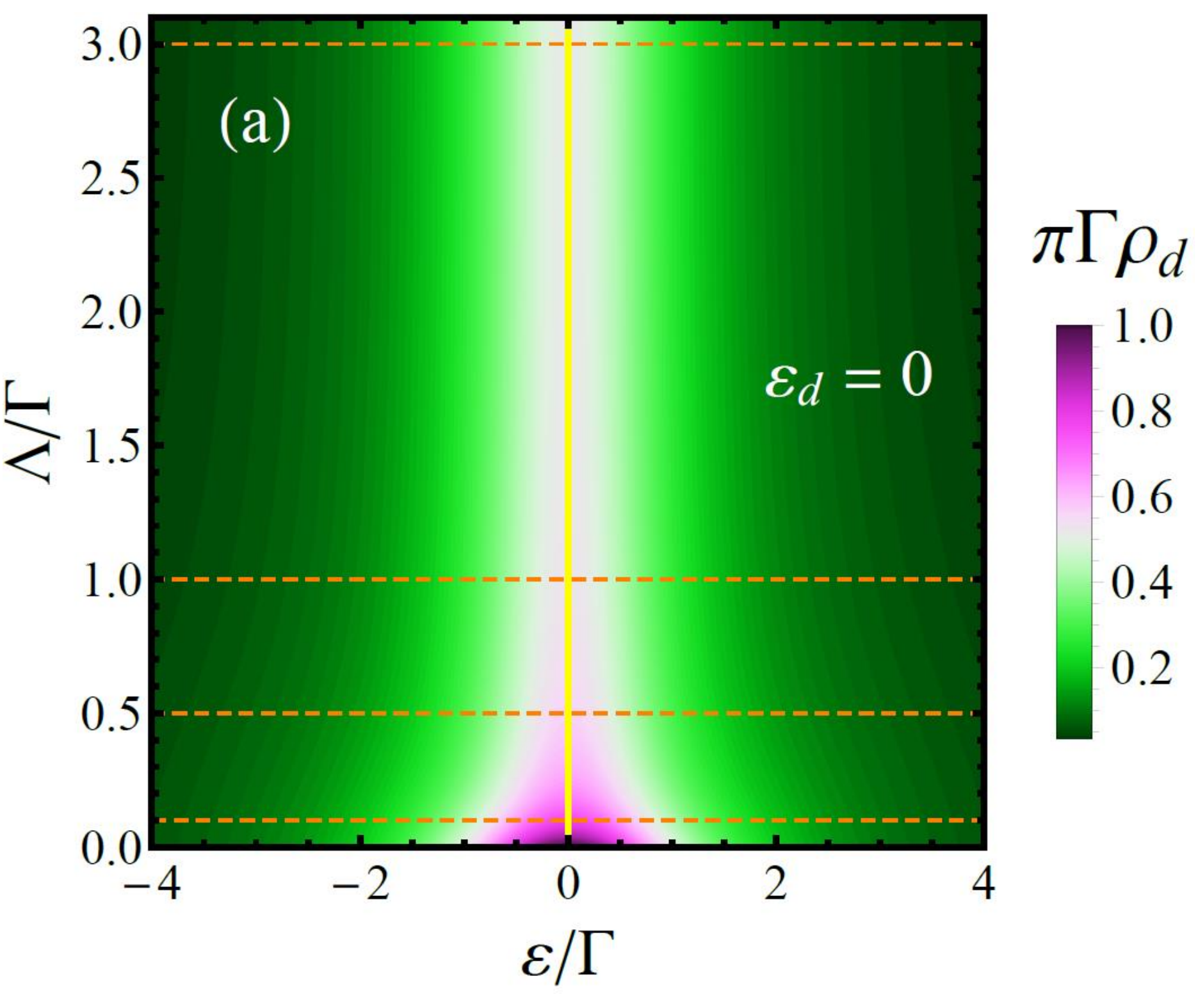}\\
\includegraphics[width=0.99\linewidth]{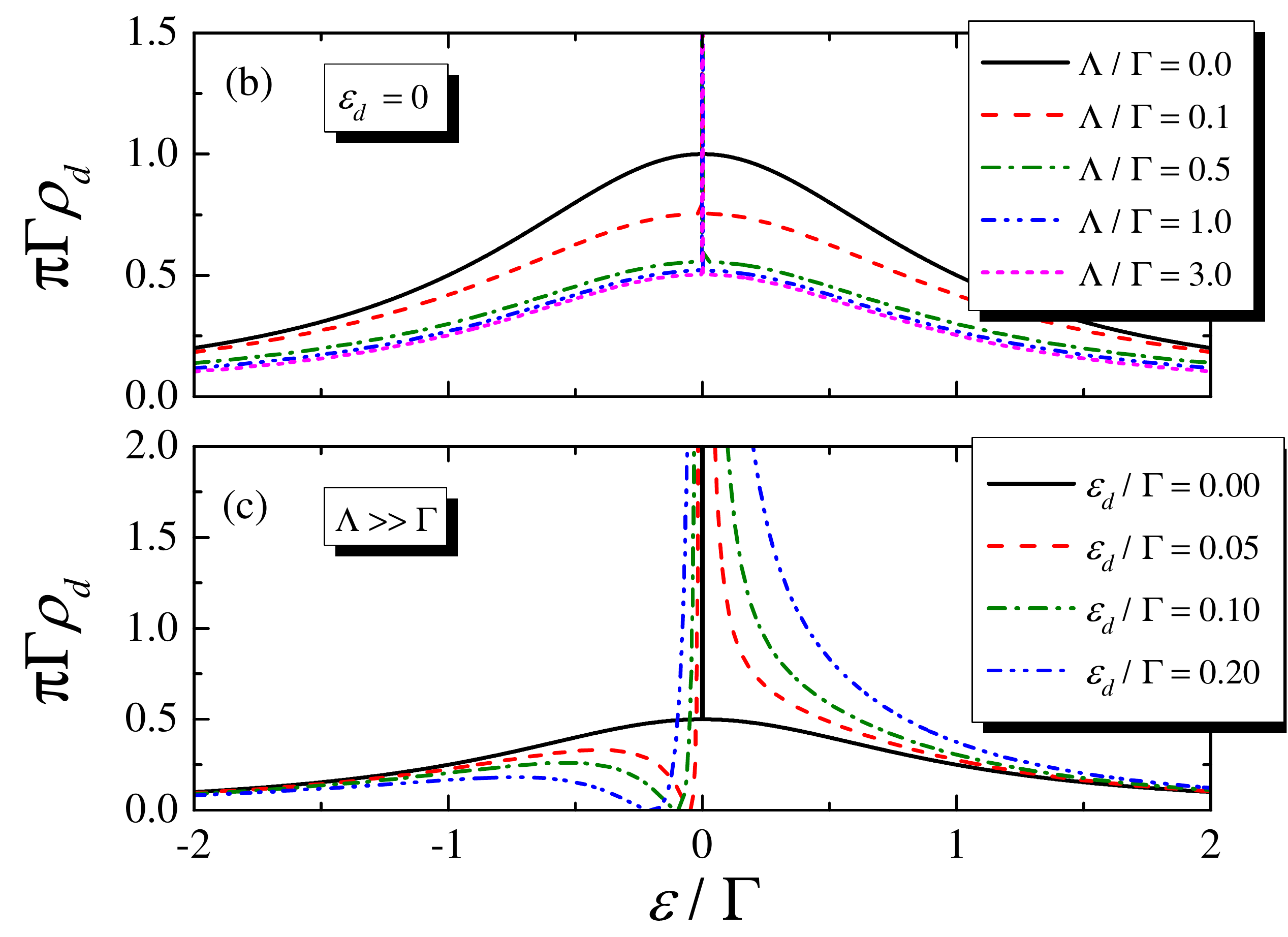}
\caption{(a) Color map of LDOS in the QD, $\rho_{d}$, as
function of the energy and $\Lambda$. The vertical solid yellow line
represents the $\delta$-Dirac function $\delta(\varepsilon)$. $\rho_{d}$ as
function of energy (b) for different $\Lambda$ values using $\ve_{d}=0$,
and (c) for different $\ve_{d}$ values using $\Lambda\gg\Gamma$. In panel
(b), the curves with $\Lambda\neq0$ correspond to the horizontal dashed
orange lines in panel (a).}
\label{fig4}
\end{figure}

The behavior of the self-energy discussed above have important
consequences in the QD LDOS, $\rho_d(\ve)$, calculated from
Eq.~\eqref{LDOS}. This quantity is the one that is actually accessible in experiment via transport spectroscopy. Figure~\ref{fig4}(a) shows LDOS as a
function of $\ve$ and $\Lambda$ for $\varepsilon_{d}=0$. For $\Lambda=0$
(uncoupled MBS) we observe  a broad peak placed around
$\varepsilon=\ve_d=0$. Once the coupling  of the MBS is turned on
($\Lambda\neq 0$), the amplitude of the LDOS decreases as $\Lambda$ increases, but the height of the peak does not go below $1/2\pi\Gamma$. Besides, at $\varepsilon=0$ sharp peak is observed. This sharp
peak is a direct consequence of the vanishing effective hybridization
function due to the presence of the MBIC. It is better appreciated in
Fig.~\ref{fig4}(b) where we show $\rho_d$ along the horizontal orange lines
of Fig.~\ref{fig4}(a). Indeed this behavior can be understood
analytically; from Eq.~\eqref{GFU0}, in the limit of strong MBS coupling
($\Lambda\gg\Gamma$), we can write
\begin{equation}
\pi\rho_{d}(\varepsilon,
\ve_{d}=0)=\frac{\Gamma}{2}\left(\frac{1}{\varepsilon^{2}+\Gamma^{2}}
\right)+\frac{\pi}{2}\delta(\varepsilon)~.
\end{equation}
Clearly, at $\ve_{d}=0$ a bound state in the continuum (BIC) is obtained
at zero energy whenever $\Lambda\neq 0$. In Fig.~\ref{fig4}(c) $\rho_{d}$
is displayed for fixed $\Lambda\gg\Gamma$ and different values of
$\ve_{d}\neq 0$. Note that the observed BIC feature evolves to a situation
with an antiresonance at $\ve=\ve_d$ for $\ve_d\neq 0$. Analytically, for
small values of $\ve_{d}$, as in Fig.~\ref{fig4}(c), we can express the
LDOS as
\begin{equation}
\pi\rho_{d}(\ve)\sim\frac{\Gamma}{2}\frac{(\ve+\ve_{d})^{2}}
{(\ve-\ve_{d})^{4} +\ve^{2}\Gamma^{2}}~.
\end{equation}
From this, we note that indeed there is an anti-resonance at $\ve=\ve_d$.
We see, therefore, that tunning $\ve_d$ is relevant to achieve a BIC. At
this point, we should emphasize that in this non-interacting scenario BIC
seen in the QD LDOS results solely from the leaking of the MBS into the continuum. In the following, we will see that this feature is still present
in the interacting regime of the QD.

\subsection{Interacting Regime}\label{subsecUno0}

\begin{figure}[]
\centering
\includegraphics[width=0.99\linewidth]{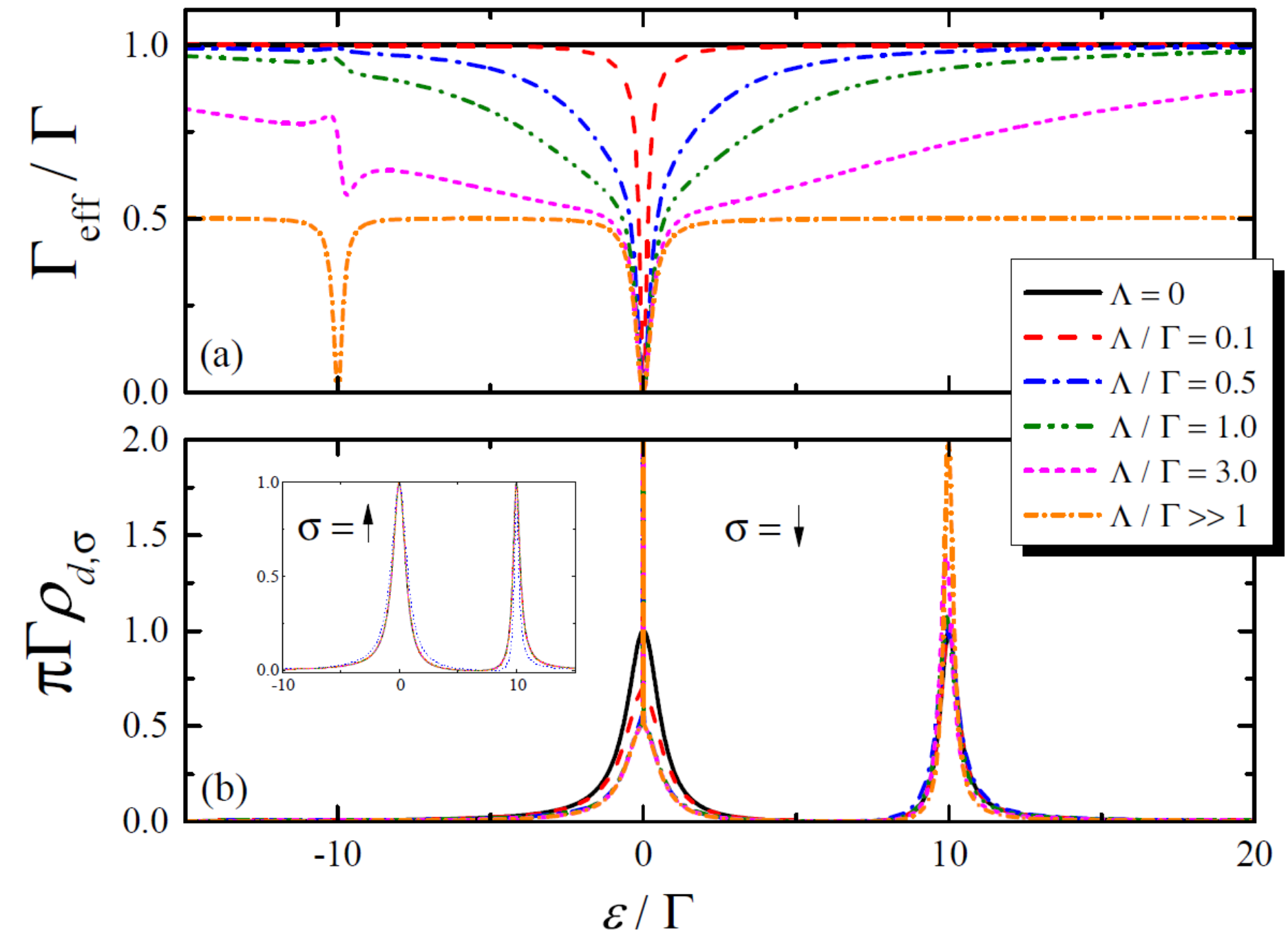}
\caption{(a) $\Gamma_{\rm eff}$ and (b) LDOS for spin $\sigma=\,\dn$ as function of the
energy for different $\Lambda$. The inset in panel (b) is the LDOS for spin $\sigma=\,\up$.}
\label{fig5}
\end{figure}

In this subsection, we study the interacting regime of the QD, $U\neq 0$.
We focus on the Coulomb blockade regime, to which the Hubbard approximation
is reasonably good. In contrast to the previous subsection, now the LDOS
depends on the temperature ($T$), and we assume $T$ larger than the
Kondo temperature $T_{K}$ so that Kondo correlations are thermally suppressed.
Again, using the equation of motion procedure in the energy domain, the GF of the QD for this case is given by
\begin{align}
&\la\la d_{\sigma};d_{\sigma}^{\dag}\ra\ra_{\ve}=\\&\frac{\ve-\ve_{d}-U(1-
\la n_{\bar{\sigma}}\ra)}{(\ve-\ve_{d})(\ve-\ve_{d}-U)-\left[\ve- \ve_{d}
-U(1-\la n_{\bar{\sigma}}\ra)\right]\Sigma_{\sigma}(\ve)}\,,\nonumber
\end{align}
where  $\Sigma_{\sigma}(\ve)$ has the form of the Eqs.~\eqref{SigmaUP} and
\eqref{SigmaDN}, with
\begin{eqnarray}
\label{MMU}
&&M(\ve)=-2i\Lambda\times\\
&&\left[\ve+\frac{2i\Lambda(\ve+\ve_{d})
(\ve+\ve_{d}+U)}{(\ve+\ve_{d})(\ve+\ve_{d}+U)+i\Gamma\left[\ve+\ve_{d}
+U(1-\la n_{\up}\ra)\right]}\right]^{-1}\nonumber.
\end{eqnarray}
Here we should emphasize that, as a consequence of the Coulomb
interaction, the GF for spin $\sigma$ depends on QD
occupation $\langle n_{\bar{\sigma}}\rangle$ given by
\begin{equation}
    \la n_{\sigma}\ra= -\frac{1}{\pi}\int_{-\infty}^{\infty}\text{d}\ve\,
\text{Im}\left[\la\la d_{\sigma};d_{\sigma}^{\dag}\ra\ra_{\ve}
\right]f(\ve).
\end{equation}
where $f(\ve)$ is the  Fermi's function. It, therefore, enforce us to
perform a self-consistent calculation numerically. To show our numerical result, we set $U=10\,\Gamma$ and carry on the numerical calculations at $k_{\rm B}T=10^{-2}\Gamma$ which happen to be above $k_{\rm B}T_{K}$ for most of the parameters used throughout this paper.
In Fig.~\ref{fig5} we show the effect of the MBS in the effective
hybridization function [Fig.~\ref{fig5}(a)] and the LDOS
[Fig.~\ref{fig5}(b)] for a fixed $\ve_{d}=0$ and various values of
$\Lambda$. Figure~\ref{fig5}(a) is similar to what displayed in
Fig.~\ref{fig2}(a) but now, for finite $U$. We observe that
$\Gamma_{\text{eff}}=\Gamma$ for $\Lambda=0$ and $\Gamma_{\text{eff}}=0$
for $\ve=\ve_{d}=0$, whenever $\Lambda\neq 0$. This behavior is very much
similar to the non-interacting case shown in Fig.~\ref{fig2}(a).  Again,
this is a direct consequence of the MBS leaking into the continuum,
reaching the physics quantities in the QD. The behavior of the curves of Fig.~\ref{fig5}(a) can be obtained analytically from
Eq.~\eqref{MMU} for large values of  $\Lambda$. In fact, for $\Lambda\gg\Gamma$,   $M(\ve)$ it is independent of
$\Lambda$. In this limit, setting $\ve_{d}=0$, we can write the effective
hybridization as
\begin{equation}
\Gamma_{\text{eff}}(\ve)=\dfrac{2\Gamma \ve^{2}(\ve+U)^{2}}{4\ve^{2}
(\ve+U)^{2} +[\ve+(1-\la n_{\up}\ra)U]^{2}\Gamma^{2}}\,.
\end{equation}
This result clearly show that $\Gamma_{\text{eff}}=0$  vanishes at both
$\ve=0$, and $\ve=-U$. Nevertheless, no important consequence in the
$\rho_{d,\sigma}$ is observed for $\ve_d=-U$ since that energy is far away from
the $\ve_d=0$
Similarly to the non-interacting case, we note also that
$\Gamma_{\text{eff}} \rightarrow\Gamma/2$ for all the energies regions such
as $|\ve|\gg \Gamma,U$.

The features observed for $\Gamma_{\text{eff}}$ are directly related to the
LDOS of the QD, which is shown in Fig.~\ref{fig5}(b). For the
case with unconnected MBS, $\Lambda=0$, two peaks are observed, of the same
amplitude, localized at energies $\ve=0$ and $\ve=U$ due to the Coulomb
blockade regime in our system. On the other hand, for the cases with
$\Lambda\neq 0$, different modifications are achieved in each of the
mentioned peaks. The amplitude of the peak located around $\ve=0$  decreases as
$\Lambda$ increases, while at exactly $\ve=0$ a very narrow peak, a BIC,
arises from the QD effective disconnection
($\Gamma_{\text{eff}}(\ve=\ve_{d}=0)=0$), is similar to the one
discussed in Sec.~\ref{subsecU0}. At this point, it is interesting to note
that the peak located at $\ve=U$, becomes narrower increasing its
amplitude, although it remains finite since $\Gamma_{\text{eff}}$ does not
vanishes. Thus, whenever the QD is in Coulomb regime, the leaked MBS into
the continuum affects the LDOS substantially, in a similar fashion as in
the non-interacting case.
\begin{figure}[h]
\centering
\includegraphics[width=0.99\linewidth]{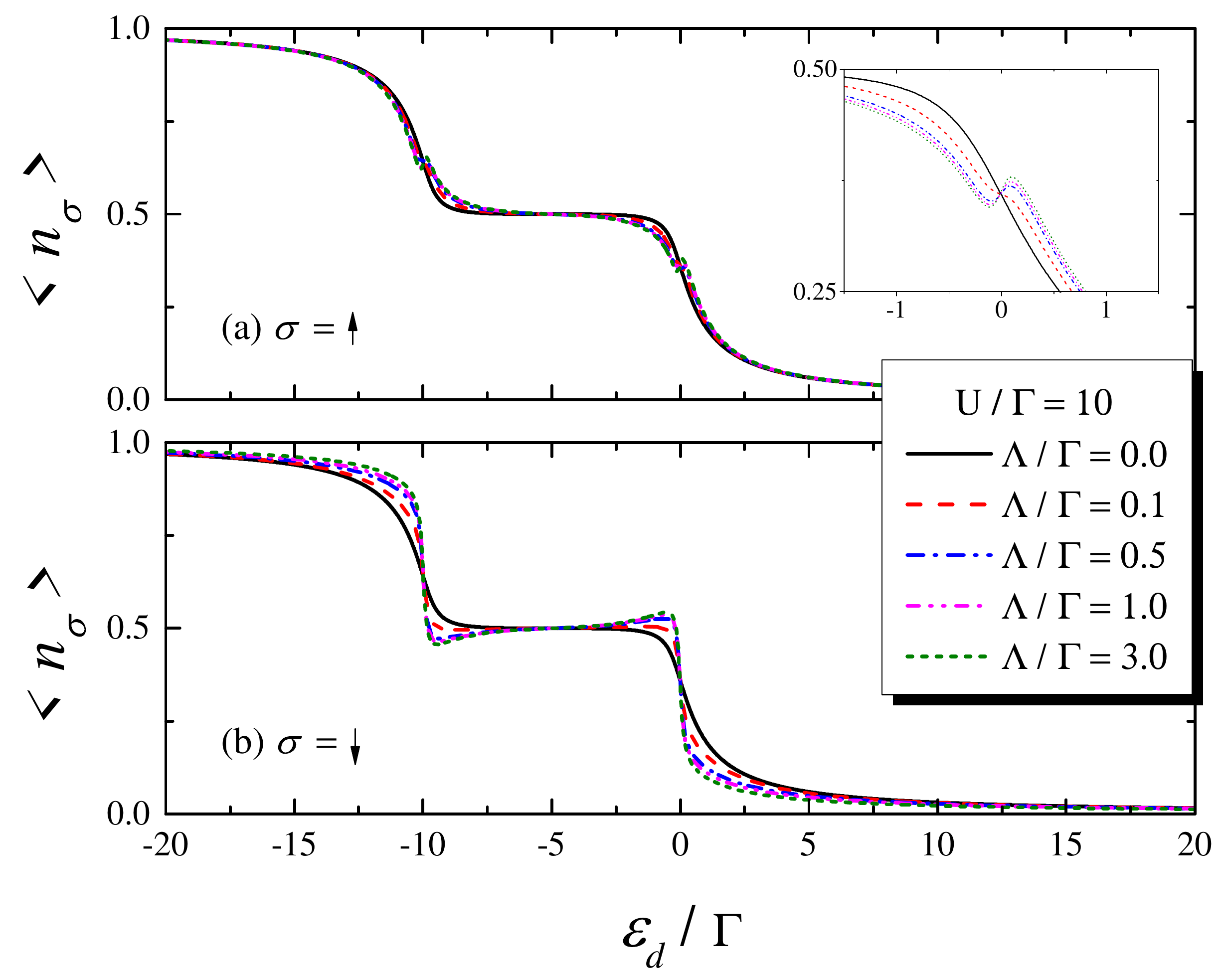}
\caption{Occupation number (a) $n_{\up}$ and (b) $n_{\downarrow}$ of the QD as function of $\ve_{d}$, for different $\Lambda$ values. The inset of panel (a) shows a zoom in of $n_\up$ around $\ve_d=0$.}
\label{fig6}
\end{figure}

Taking  into account the discussion above, the QD local density of  the QD
for spin \emph{down} can be written as
\begin{widetext}
\begin{equation}
    \pi\rho_{d\dn}(\ve)= \frac{\Gamma}{2}\frac{(\ve+U)^{2}(\ve-(1-\la
n_{\up}\ra)U)^{2}}{\ve^{6}+(1-\la n_{\up}\ra)^{2}U^{4}\Gamma^{2}+\ve^{4}
(\Gamma^{2}-2U^{2})+\ve^{2}U^{2}(U^{2 }-2(1-\la
n_{\up}\ra)\Gamma^{2})}+\frac{\pi}{3}\delta(\ve)\,.
\end{equation}
\end{widetext}
Despite the complexity of the equation  above, for the corresponding
occupancy, we can extract that the wide peak placed around $\ve=0$
asymptotically reach $\pi\rho_{d,\dn}\rightarrow 1/2\Gamma$, while the one
located at $\ve=U$ increase up to $\pi\rho_{d,\dn}\rightarrow 2/\Gamma$,
both in the limit $\Lambda\gg\Gamma$.

Before closing this section, we show how the presence of the MBIC affects
the occupation of the QD. In Fig.~\ref{fig6}, we show the spin-resolved occupation number in the QD as a function of $\ve_{d}$. From this figure, for
$\Lambda=0$ (solid black lines) there is spin degeneration in the
occupancy, as we expected since the Hamiltonian is spin symmetric for this case. Allowing coupling between the MBS and the continuum,
$\Lambda\neq 0$, the spin symmetry brokes, and deviations are observed.
As consequence, in Fig.~\ref{fig6}(a) we observe a subtle oscillation of
$n_{\up}$ around $\ve_d=0$ and $\ve_d=-U$, better seen in the inset for
energies near $\ve_d=0$. In Fig.~\ref{fig6}(b) we show the corresponding
curves for $n_{\dn}$. Here, a more interesting consequence of the MBS
visible. Note that, while for $\Lambda=0$ the occupancy always increase as
we decrease $\ve_d$ (same happening to $\langle n_\up\rangle$), for finite
$\Lambda$ $\langle n_\dn\rangle$ decreases with $\ve_d$ within the interval
$-U<\ve_d<0$. By noting that for a given $\ve_d$ in this interval $\langle
n_\dn\rangle$ decreases while $\langle n_\up\rangle$ increases as $\Lambda$ increases, we conclude that there is a spin polarization in the QD. It
can  interpret as an effective magnetic field due to the presence of the
MBS that breaks time-reversal symmetry of the system.

\section{Conclusions}\label{Conclu}

We studied a system formed by a QD coupled to the continuum, which is connected to an MBS localized at the end of a TSW. Considering that
continuum electrons with a particular spin \emph{down} couples with the
MBS, we found that the leakage of the MBS into the continuum affects the physical properties of the QD greatly. As a consequence of this leaking, the
QD becomes effectively decoupled from the rest of the system at energies
$\ve=\ve_{d}=0$, for both cases, interacting and non-interacting regimes.
In the interacting case, the second peak due to Coulomb blockade,
placed at $\ve=\ve_{d}+U$, is also affected by the MBIC. It becomes narrower and increasing its amplitude as the coupling strength between MBS and continuum increases. Besides, we have performed an analytic treatment of
the effective coupling and local density of states in the limit of strong
MBS-continuum coupling. Owing to the robustness of the MBS against the applied
electric field, MBICs provide an exciting manner to control the QDs
electronic properties without changing the energy position of the bound
state in the continuum.\\

\acknowledgments
J.P.R.-A is grateful for the funding of FONDECYT Postdoc. Grant No. 3190301 (2019). P.A.O. acknowledges support from FONDECYT Grant No. 1180914. E.V. thanks the brazilian agencies CAPES, CNPq and FAPEMIG for support.

\onecolumngrid
\appendix
\section{QD Green's function}\label{appa}

In this appendix, we show the procedure used to reach an analytic expression
for the QD retarded Green's function in our system. We considered the
equation of motion method up to the equations hierarchy that allows describing the Coulomb blockade phenomena in the QD. The system Hamiltonian
is given by Eq.\ (\ref{H}). Note that it is not symmetrical in spin degree
of freedom, since only continuum electrons with spin $\sigma=\,\dn$ are
coupled with the MBS [Eq.\ (\ref{HcMBS})]. The general expression for the
retarded Green's function equation of motion in the energy domain is given by
\begin{equation}\label{EOMegeneral}
(\ve+i0^{+})\la\la A;B\ra\ra_{\ve}^{r}=\la \{A;B\}\ra+\la\la [A;H];B\ra\ra_{\ve}^{r}\,,
\end{equation}
where $A$ and $B$ are two arbitrary operators, and $0^{+}$ an
infinitesimal (positive) number. Throughout this section, as in the main
text, we display the energy as $\ve+i0^{+}\rightarrow\ve$ for simplicity.

Using Eq. \eqref{EOMegeneral}, for spin $\sigma=\,\dn$ electrons,
calculating the corresponding commutators/anticonmutators, the first
hierarchy of equations are
\begin{equation}\label{ddM1}
(\ve-\ve_{d})\la\la d_{\dn};d_{\dn}^{\dag}\ra\ra_{\ve}=1+\sum_{\bf{k}}
V\la\la c_{\bf{k},\dn};d_{\dn}^{\dag}\ra\ra_{\ve}+ U\la\la
n_{\up}d_{\dn};d_{\dn}^{\dag}\ra\ra_{\ve}\,
\end{equation}
\begin{equation}
\label{cdM1}
(\ve-\ve_{\bf{k}})\la\la c_{\bf{k},\dn};d_{\dn}^{\dag}\ra\ra_{\ve} =V\la\la
d_{\dn};d_{\dn}^{\dag}\ra\ra_{\ve}-\lambda\la\la\gamma_{1};d_{\dn}^{\dag}
\ra\ra_{\ve}\,,
\end{equation}
\begin{equation}
\label{Md}
\ve\la\la\gamma_{1};d_{\dn}^{\dag}\ra\ra_{\ve}=-2\lambda\sum_{\bf{k}'}
\left(\la\la c_{\bf{k}',\dn};d_{\dn}^{\dag}\ra\ra_{\ve}-\la\la
c_{\bf{k}',\dn}^{\dag};d_{\dn}^{\dag}\ra\ra_{\ve}\right)\,,
\end{equation}
\begin{equation}
\label{cdM1ano}(\ve+\ve_{\bf{k}})\la\la
c_{\bf{k},\dn}^{\dag};d_{\dn}^{\dag} \ra\ra_{\ve}=-V\la\la
d_{\dn}^{\dag};d_{\dn}^{\dag}\ra\ra_{\ve}+\lambda\la\la\gamma_{1};d_{\dn}^{
\dag}\ra\ra_{\ve}\,,
\end{equation}
\begin{equation}
\label{ddM1ano}
(\ve+\ve_{d})\la\la d_{\dn}^{\dag};d_{\dn}^{\dag}\ra\ra_{\ve}=-
\sum_{\bf{k}}V\la\la  c_{\bf{k},\dn}^{\dag};d_{\dn}^{\dag}\ra\ra_{\ve}
-U\la\la n_{\up}d_{\dn}^{\dag};d_{\dn}^{\dag}\ra\ra_{\ve}\,,
\end{equation}
where we have suppressed the superscript $r$ for simplicity. As a consequence of MBS presence, anomalous Green's function that must be calculated. The next hierarchy of equations is extracted from the last
terms in Eqs.\ (\ref{ddM1}) and (\ref{ddM1ano}). They lead to
\begin{equation}\label{HastaMHubbard}
(\ve-\ve_{d}-U)\la\la n_{\up} d_{\dn};d_{\dn}^{\dag}\ra\ra_{\ve}=\la
n_{\up} \ra+\sum_{\bf{k}}V\la\la n_{\up}c_{\bf{k},\dn};d_{\dn}^{\dag}
\ra\ra_{\ve} +\sum_{\bf{k}}V\la\la
d_{\up}^{\dag}c_{\bf{k},\up}d_{\dn};d_{\dn}^{\dag} \ra\ra_{\ve}
-\sum_{\bf{k}}V\la\la c_{\bf{k},\up}^{\dag}d_{\up}d_{\dn};d_{\dn}^{\dag}\ra\ra_{\ve}\,,
\end{equation}
\begin{equation}
(\ve+\ve_{d}+U)\la\la n_{\up} d_{\dn}^{\dag};d_{\dn}^{\dag}\ra\ra_{\ve}
=-\sum_{\bf{k}}V\la\la n_{\up}c_{\bf{k},\dn}^{\dag};d_{\dn}^{\dag}
\ra\ra_{\ve}+\sum_{\bf{k}} V\la\la d_{\up}^{\dag}c_{\bf{k},\up}
d_{\dn}^{\dag} ;d_{\dn}^{\dag}\ra\ra_{\ve} -\sum_{\bf{k}}V\la\la
c_{\bf{k},\up}^{\dag} d_{\up}d_{\dn}^{\dag};d_{\dn}^{\dag}\ra\ra_{\ve}\,.
\label{HastaMHubbardano}
\end{equation}
Up  to this point, it is possible to reach the Coulomb blockade regime.
Employing the Hubbard approximation, Eqs.\ (\ref{HastaMHubbard}) and
(\ref{HastaMHubbardano}) can be seen as follows
\begin{eqnarray}\label{HastaMHubbardapro}
(\ve- \ve_{d}-U)\la\la n_{\up} d_{\dn};d_{\dn}^{\dag}\ra\ra_{\ve}&=&\la
n_{\up}\ra+\sum_{\bf{k}}V\la n_{\up}\ra \la\la
c_{\bf{k},\dn};d_{\dn}^{\dag}\ra\ra_{\ve}\,,\\
\label{HastaMHubbardanoapro}
(\ve+\ve_{d}+U)\la\la n_{\up}
d_{\dn}^{\dag};d_{\dn}^{\dag}\ra\ra_{\ve}&=&-\sum_{\bf{k}}V\la n_{\up}\ra
\la\la c_{\bf{k},\dn}^{\dag};d_{\dn}^{\dag}\ra\ra_{\ve}\,,
\end{eqnarray}
where we have considered $\sum_{\bf{k}}\la
d_{\sigma}^{\dag}c_{\bf{k},\sigma}\ra=\sum_{\bf{k}}\la
c_{\bf{k},\sigma}^{\dag}d_{\sigma}\ra$. Replacing Eq.\
(\ref{HastaMHubbardanoapro}) into Eq.\ (\ref{ddM1ano}), we have
\begin{equation}
(\ve+\ve_{d})\la\la
d_{\dn}^{\dag};d_{\dn}^{\dag}\ra\ra_{\ve}=-\left(1-\frac{U\la
n_{\up}\ra}{\ve+\ve_{d}+U}\right)\sum_{\bf{k}}V\la\la
c_{\bf{k},\dn}^{\dag};d_{\dn}^{\dag}\ra\ra_{\ve}\,,
\end{equation}
thus, including this result into Eq.\ (\ref{cdM1ano}), we obtain
\begin{equation}
\left[1-\frac{V^{2}\tilde{g}(\ve)}{\ve+\ve_{d}}\left(1-\frac{U \la
n_{\up}\ra}{\ve+\ve_{d}+U}\right)\right]\sum_{{\bf k}}\la\la
c_{\bf{k},\dn}^{\dag};d_{\dn}^{\dag}\ra\ra_{\ve}=\lambda\tilde{g}
(\ve)\la\la\gamma_{1};d_{\dn}^{\dag}\ra\ra_{\ve}\,,
\end{equation}
where  we have defined
$\tilde{g}(\ve)=\sum_{\bf{k}}(\ve+\ve_{\bf{k}})^{-1}$. Then, the Eq.\
(\ref{Md}) is rewritten as
\begin{equation}
\left(\ve-2\lambda^{2}\tilde{g}
(\ve)\left[1-\frac{V^{2}\tilde{g}(\ve)}{\ve+\ve_{d}}\left(1-\frac{U\la
n_{\up}\ra}{\ve+\ve_{d}+U}\right)\right]^{-1}\right)\la\la\gamma_{1};d_{\dn
}^{\dag}\ra\ra_{\ve}=-2\lambda\sum_{\bf{k}}\la\la
c_{\bf{k},\dn};d_{\dn}^{\dag}\ra\ra_{\ve}\,.
\end{equation}
Consequently, the Eq.\ (\ref{cdM1}) is expressed as
\begin{equation}
\left(1-2\lambda^{2}g(\ve)\left[\ve-\frac{2\lambda^{2}\tilde{g}
(\ve)}{1-\dfrac{V^{2}\tilde{g}(\ve)}{\ve+\ve_{d}}\left(1-\dfrac{U\la
n_{\up}\ra}{\ve+\ve_{d}+U}\right)}\right]^{-1}\right)\sum_{\bf{k}}\la\la
c_{\bf{k},\dn};d_{\dn}^{\dag}\ra\ra_{\ve}=Vg(\ve)\la\la
d_{\dn};d_{\dn}^{\dag}\ra\ra_{\ve}\,,
\end{equation}
being defined $g(\ve)=\sum_{\bf{k}}(\ve-\ve_{\bf{k}})^{-1}$. On the other
hand, after replacing the Eq.\ (\ref{HastaMHubbardapro}) into Eq.\
(\ref{ddM1}) we obtain
\begin{equation}
(\ve-\ve_{d})\la\la d_{\dn};d_{\dn}^{\dag}\ra\ra_{\ve}=1+ \frac{U\la
n_{\up}\ra}{\ve-\ve_{d}-U}+\left(1+\frac{U\la
n_{\up}\ra}{\ve-\ve_{d}-U}\right)\sum_{\bf{k}}V\la\la
c_{\bf{k},\dn};d_{\dn}\ra\ra_{\ve}\,,
\end{equation}
which allow a closed solution for the set of equations. Finally,
performing algebraic manipulations we have
\begin{equation}\label{ddcasifinal}
\la\la d_{\dn};d_{\dn}^{\dag}\ra\ra_{\ve}=\dfrac{\ve-\ve_{d}-U(1-\la n_{\up}\ra)}{(\ve-\ve_{d})(\ve-\ve_{d}-U)-\dfrac{(\ve-\ve_{d}-U(1-\la n_{\up}\ra))V^{2}g(\ve)}{1-M(\ve)}}\,,
\end{equation}
where
\begin{equation}\label{Mcasifinal}
M(\ve)=2\lambda^{2}g(\ve)\left[\ve-\frac{2\lambda^{2}
\tilde{g}(\ve)(\ve+\ve_{d})(\ve+\ve_{d}+U)}{(\ve+\ve_{d})(\ve+\ve_{d}+U)-V^
{2}\tilde{g}(\ve)(\ve+\ve_{d}+U(1-\la n_{\up}\ra))}\right]^{-1}\,.
\end{equation}
At this point, it is interesting to note that the quantities $V^{2}g(\ve)$
and $\lambda^{2}g(\ve)$, can be treated within the wideband approximation.
In this limit, they are energy independent and fulfill electron-hole
symmetry, such as
\begin{eqnarray}
V^{2}g(\ve)=V^{2}\tilde{g}(\ve)=-i\Gamma\,,\\
\lambda^{2}g(\ve)=\lambda^{2}\tilde{g}(\ve)=-i\Lambda\,.
\end{eqnarray}
Then, the QD Green's function for $\sigma=\,\dn$ is given by
\begin{equation}\label{ddfinal}
\la\la d_{\dn},d_{\dn}^{\dag}\ra\ra_{\ve}=\frac{\ve-\ve_{d}-U(1-\la
n_{\up}\ra)}{(\ve-\ve_{d})(\ve-\ve_{d}-U)-(\ve-\ve_{d}-U(1-\la
n_{\up}\ra))\Sigma_{\dn}(\ve)}\,,
\end{equation}
where  $\Sigma_{\dn}(\ve)=-i\Gamma/[1-M(\ve)]$ and all the MBS contribution
is embedded in the function
\begin{equation}
M(\ve)=
-2i\Lambda\left[\ve+\frac{2i\Lambda(\ve+\ve_{d})(\ve+\ve_{d}+U)}{(\ve+\ve_{
d})(\ve+\ve_{d}+U)+i\Gamma(\ve+\ve_{d}+U(1-\la n_{\up}\ra))}\right]^{-1}\,.
\end{equation}

For  the component $\sigma=\,\up$, we note that up to the hierarchy
considered in this paper, there is no MBS explicit contribution in the
corresponding Green's function. Therefore, it can be obtained from Eq.\
(\ref{ddfinal}) fixing $\lambda=\Lambda=M(\ve)=0$, then
$\Sigma_{\up}(\ve)=-i\Gamma$ and
\begin{equation}
\la\la d_{\up};d_{\up}^{\dag}\ra\ra_{\ve}=\frac{\ve-\ve_{d}-U(1-\la
n_{\dn}\ra)}{(\ve-\ve_{d})(\ve-\ve_{d}-U)-(\ve-\ve_{d}-U(1-\la
n_{\dn}\ra))\Sigma_{\up}(\ve)}\,.
\end{equation}

\twocolumngrid

\bibliography{biblio}
\bibliographystyle{apsrev4-1}

\end{document}